\documentclass{sig-alternate}

\usepackage{url}
\usepackage{subfigure,graphicx}
\usepackage[noend]{algpseudocode}
\usepackage{amsmath,amssymb}
\usepackage{multirow}

\usepackage{booktabs}

\newcommand{\Yelp}{\texttt{Yelp}}
\newcommand{\YelpUnaware}{\texttt{YelpPre}}
\newcommand{\YelpAware}{\texttt{YelpPost}}
\newcommand{\YelpUnbiased}{\texttt{YelpUnbiased}}

\begin{document}

\title{The warm-start bias of Yelp ratings}

\numberofauthors{1} 

\author{
\alignauthor
Michalis Potamias\\
\affaddr{Smart Deals}\\ Groupon\\ 
\email{michalis@groupon.com}\titlenote{Data used in this study have been collected from the web, while the author was a PhD candidate at Boston University.}
}

\maketitle
\begin{abstract}
Yelp ratings are often viewed as a reputation metric for local businesses. In this paper we study how Yelp ratings evolve over time. Our main finding is that on average the first ratings that businesses receive overestimate their eventual reputation. In particular, the first review that a business receives in our dataset averages 4.1 stars, while the 20th review averages just 3.69 stars. This significant warm-start bias which may be attributed to the limited exposure of a business in its first steps may mask analysis performed on ratings and reputational ramifications. Therefore, we study techniques to identify and correct for this bias. Further, we perform a case study to explore the effect of a Groupon deal on the merchant's subsequent ratings and show both that previous research has overestimated Groupon's effect to merchants' reputation and that average ratings anticorrelate with the number of reviews received. Our analysis points to the importance of identifying and removing biases from Yelp reviews.
\end{abstract}

\section{Introduction} \label{sec:introduction}

The past few years the web has become an emerging marketing channel for local businesses, such as restaurants, beauty salons, auto services, etc. In particular, this channel is powered by the broad popularity of review sites such as Yelp and Angieslist, reservation sites such as Opentable, social networking sites that allow their users to check-in such as Facebook and Foursquare, and discount offer sites such as Groupon and Livingsocial.

Yelp is a  website that hosts users reviews about local businesses. Users can write a review and rate a business with 1,2,3,4, or 5 stars. Posting a review in Yelp is easy; the sole requirement is a valid email address. Yelp also offers search functionality: users can search for local businesses using terms and locations. Yelp responds with relevant search results displaying the average review rounded in half stars (e.g., 3.5, 4 stars etc.) and the number of reviews that the business has received. The average rating may be viewed as a metric for the quality of the business while the number of reviews give confidence to that value. Anonymous users can introduce a Yelp rating and the same users can review a business multiple times, making Yelp's number of reviews and average review score susceptible to manipulations and biases. This has been a source of criticism for Yelp~\cite{regression, regulation, mechturkyelp}.

In this paper we study data collected via Yelp's API. Our main finding is that Yelp average ratings suffer from a bias observable mainly in the initial reviews. To identify the bias we average the ratings on the order in which they were received. We find that even though the first reviews received by merchants average 4.1 stars and the second review averages 4.0 stars, eventually the average rating drops to 3.7 stars after the 20th review. This bias can be attributed to the exposure of a business to a limited population during its first steps. 

In the second part of this paper we demonstrate the effect of the bias via a case study. We combine the Yelp dataset with data collected from Groupon. Groupon is a large daily-deal website powering local business discovery via discount offers. All of the businesses in our dataset have ran a Groupon deal at some point in the past. We study ratings both before and after a business runs a Groupon deal. Not surprisingly, the wide exposure that Groupon brings to a merchant cause the frequency of Yelp reviews to almost double the month after the Groupon offer~\cite{Byers2011}. In our case study, we analyze Groupon's effect on a merchant's reputation with and without the bias. Our main finding is that the warm-start bias leads to overestimating post-Groupon reputational ramifications~\cite{Byers2011}. Moreover we observe that ratings submitted after a business has ran a Groupon deal do not suffer from the bias indicating that these reviews are more likely to be real.

The remainder of this paper is organized as follows. In Section~\ref{sec:related} we present published research related to our work. In Section~\ref{sec:data} we describe the data used in this paper. In Section~\ref{sec:yelp} we identify and correct for the bias of the initial reviews in Yelp and in Section~\ref{sec:groupon} we demonstrate the severity of this bias in the Groupon case study. We conclude the paper in Section~\ref{sec:conclusion}.

\section{Related work}
\label{sec:related}

Online ratings have become a part of local commerce lifecycle. In a recent study Luca~\cite{Luca2011} concludes that consumer review websites improve the information available about the product quality of restaurants. He argues that a one-star rating increase contributes to a 5-9\% increase in revenue and that a change in a restaurant's rating has 50\% more impact when a restaurant has at least 50 reviews. The author suggests that both the number of reviews and the average rating contribute to revenue for local businesses. The details of this interplay comprise an interesting future research direction. The study does not consider the ranking mechanism of Yelp; it is possible that Yelp's popular search functionality and its implicit ranking mechanism affects the revenue as well.

Yelp's open policy has been criticized as easy to game. Anonymous users can introduce a Yelp rating and the same users can review a business multiple times, making Yelp's number of reviews and average review score susceptible to manipulation. Examples of such gaming using paid workers on Amazon's Mechanical Turk and other similar platforms have found their way into popular media~\cite{mechturkyelp}. In their study Anderson and Magruder conclude that merchants are incentivized to manipulate Yelp ratings~\cite{regression}. Further, legal considerations regarding online ratings are discussed by Goldman~\cite{regulation}. 

In this paper we discuss the notions of  ``average'' and ``number of reviews'' computed using information made publicly available by Yelp. Yelp displays a number of reviews and an average rating next to each merchant in its search response. These may differ from or coincide with the ones we calculate with the publicly available data. Also, in this paper we do not study how Yelp ranks merchants given a location, search terms, data about merchants, and possibly other information. We also do not study Yelp's techniques to protect its content from fraudulent reviews.

Daily deals research was pioneered by our study~\cite{BMPZ2011}. The paper presents analysis of daily deals, numerous insights with regard to scheduling practices and the price elasticity of coupons and models to predict the sales of a deal. In follow-up work, Byers et al. cross Groupon with Yelp to extract more insights about daily deals diffusion and reputational ramifications~\cite{Byers2011}. The authors present a case study of Groupon's effect on the reputation of a business. In particular, they observe that the frequency that a local merchant receives reviews increases by 80\% the month after the Groupon deal. They also suggest that the rating drops by 0.12 stars following the Groupon deal. This claim was covered by popular media~\cite{mitbyers, nytimesbyers}. In the second part of this paper we revisit this case study and show that this conclusion is mainly a result of the warm-start bias of Yelp ratings.

More studies on daily deals sites have appeared that are orthogonal to our work.  Edelman et al. model the Return of Investment (ROI) of daily deal sites~\cite{Edelman2010To}. The tipping point of Groupon deals has also received attention recently~\cite{Ye2011, Grabchak2011}. Finally, Dholakia surveys 150 merchants to find that two out of three of these merchants deem their Groupon deal profitable for their business~\cite{Dholakia2010How} .

\begin{figure}[t]
 \begin{center}
     \includegraphics[width=8.0 cm]{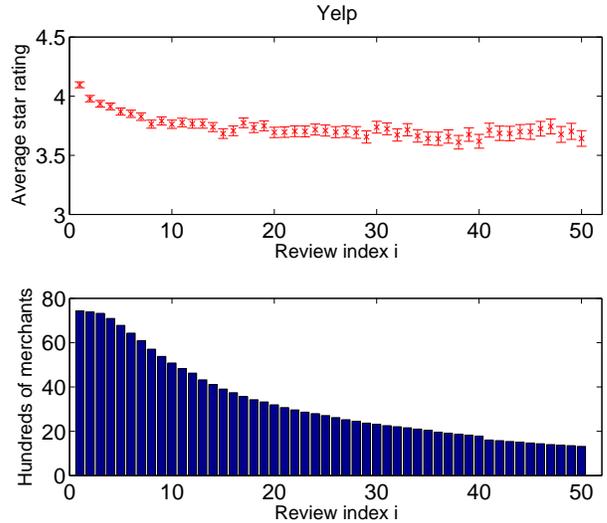}
 \end{center}
 \caption{{\small {\Yelp} dataset: Yelp Ratings by rating index. The upper plot shows the average rating at index $i$ and the corresponding 95\% confidence intervals. The lower plot shows the number of reviews available at each index. \label{fig:indexReview}}}
\end{figure}

\begin{figure}[t]
 \begin{center}
     \includegraphics[width=8.0 cm]{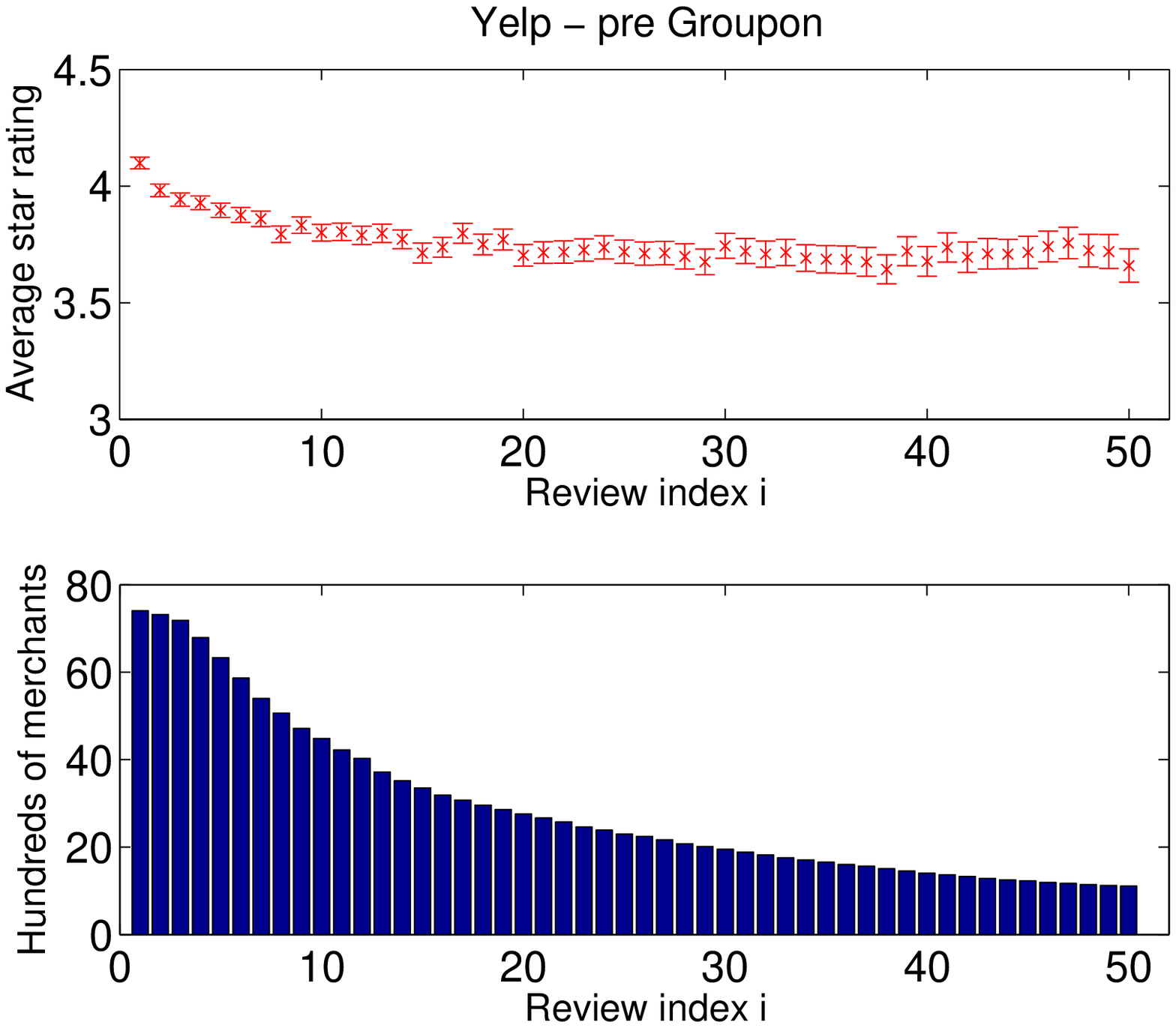}
 \end{center}
 \caption{{\small {\YelpUnaware} dataset (pre-Groupon): Yelp Ratings by rating index. The upper plot shows the average rating at index $i$ and the corresponding 95\% confidence intervals. The lower plot shows the number of reviews available at each index. \label{fig:indexReviewPre}}}
\end{figure} 

\begin{figure}[t]
 \begin{center}
     \includegraphics[width=8.0 cm]{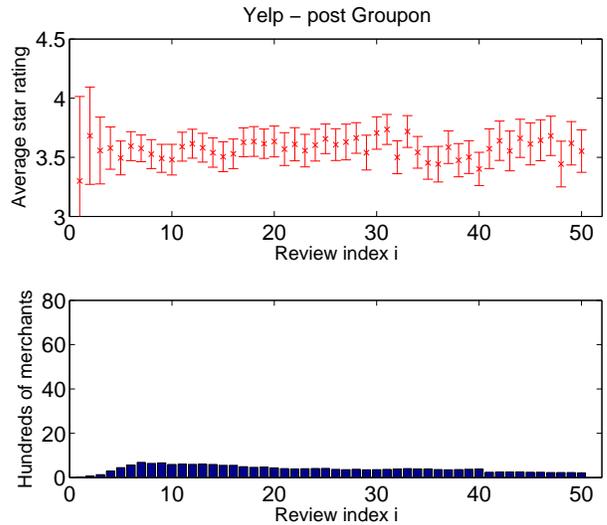}
 \end{center}
 \caption{{\small {\YelpAware} dataset (post-Groupon): Yelp Ratings by rating index. The upper plot shows the average rating at index $i$ and the corresponding 95\% confidence intervals. The lower plot shows the number of reviews available at each index. \label{fig:indexReviewPost}}}
\end{figure}

\section{Data}
\label{sec:data}

The dataset we use in this paper consists of two parts. The first part has been obtained using the Yelp API. It consists of 274344 reviews for 7426 merchants who ran at least one Groupon deal during the observation period. The reviews in our dataset have been submitted from January 2005 to August 2011. The dataset is incomplete, namely on average 23\% of reviews are not reported from Yelp's API~\cite{Byers2011}. The second part was collected from Groupon's website for the same merchants.  The deals ran in Groupon from April 2010 to July 2011. Description and analysis of this dataset is publicly available~\cite{BMPZ2011, Byers2011}.  

Each record in our dataset summarizes a review action. In particular, it consists of the merchant that this review refers to, the date that the review was submitted, and its star rating. The star rating can be 1,2,3,4, or 5 stars -- note though that Yelp reports averages using half stars. The Groupon part of the data consists of one record per merchant that appears in the Yelp part. Each record corresponds to a Groupon deal and the date that the deal launched. We refer to the complete dataset as {\Yelp}.

\section{Yelp: The initial reviews bias}\label{sec:yelp}

In this part we present the main finding of this paper. In a nutshell, the initial reviews that a merchant receives significantly overestimate the merchant's eventual reputation.

We order each merchant's ratings according to the time they were posted. Then we assign an \emph{index} $i$ to each review. Subsequently, we take the average star rating of each review index across all merchants in our dataset. We plot the result in Figure~\ref{fig:indexReview}. We limit this figure to index 50 noting that only 18\% of the merchants in the dataset have more than 50 reviews. We also plot the number of merchants with at least $i$ reviews in the histogram of the lower subplot. We note that that the 95\% confidence intervals for the mean become looser as the index increases; indicatively, even though all merchants in our dataset have at least one review, only half of the businesses have at least 16 reviews.

Observe that the 1st rating averages 4.1 stars while the 20th rating has an average of just 3.69 stars, 0.41 stars lower. This observation is not limited to the first index. Ratings with indices in $\left[1, 5\right]$ average 3.96 stars while ratings with indices in $\left[21, 25\right]$  average 3.7 stars. Similarly ratings in indices $\left[1, 20\right]$ average 3.83 stars, ratings in indices $\left[21, 40\right]$ average 3.69 stars, same as ratings in the index interval $\left[41, 60\right]$. Ratings stabilize after the 20th index.

The bias of the first 20 ratings has a strong presence in this dataset due to the long tail of the distribution of the number of ratings per merchant  as depicted in the histogram of Figure~\ref{fig:indexReview}. Indicatively we report that 38\% of the rating records  in our dataset are in the ``bias area'', i.e., they have an index smaller than 20. Also, the majority (57\%) of the merchants have less than 20 reviews.

\smallskip
\noindent
\textbf{Groupon effect. }
All merchants in {\Yelp} launched a Groupon deal at some point in time. Therefore, the bias described above could be correlated to the Groupon deal. We study if this is the case by deriving two datasets from {\Yelp}. We call the first dataset {\YelpUnaware}: This is the dataset that consists of the selection of all reviews in {\Yelp} posted before the launch date of the merchant's Groupon deal. We also derive the {\YelpAware} dataset as follows. From {\Yelp} we select all ratings that were posted \emph{after} the launch date of the Groupon deal. We redraw the plots that we drew for the {\Yelp} dataset using the  two derived datasets and we maintain the same ranges on all axes for all three plots.

Figure~\ref{fig:indexReviewPre} presents the average index review in {\YelpUnaware}. The bias found in {\Yelp} is still present and all quantitative observations from above hold. We conclude that the initial review bias is an intrinsic characteristic of Yelp reviews. Presumably, it is the result of a new business' exposure to a limited audience. This narrow audience overestimates the eventual reputation of a merchant.

Figure~\ref{fig:indexReviewPost} presents the average index review in {\YelpAware}.  Interestingly, the {\YelpAware} figure does not suffer from this bias.  Furthermore, the ratings in {\YelpUnaware} converge to the ratings of {\YelpAware} after the 20th index. The latter observation suggests that Groupon's reach to a wide audience results to ratings that match the eventual reputation of a merchant, bypassing the bias period observed in  {\YelpUnaware}.

\smallskip
\noindent
\textbf{Bias correction. }
Numerous bias correction techniques can be found in the literature. We correct this bias by applying a cutoff.  For the remainder of this paper we remove all reviews indexed in the $\left[1,20\right]$ index interval. We note that the technique of removing bias by applying a cutoff is controlled by a tradeoff. Increasing the cutoff value may remove more biased records but it also limits the data and decreases the statistical significance of results.

\begin{figure}[t]
 \begin{center}
     \includegraphics[width=8.0 cm]{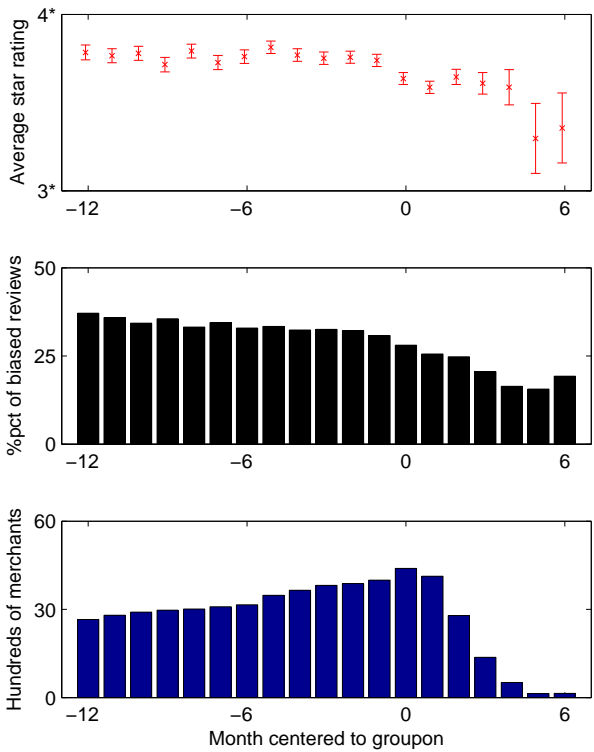}
 \end{center}
 \caption{{\small {\Yelp} dataset: Yelp Ratings by rating index. The upper plot shows the mean of the average rating of every merchant for every month around the date that a merchant ran a Groupon deal (month 0) and the corresponding 95\% confidence intervals.  The middle plot shows the percentage of biased reviews out of all reviews aggregated for each month. The lower plot shows the number of merchants aggregated at each month. \label{fig:yelp-biased-hist}}}
\end{figure}

\begin{figure}[t]
 \begin{center}
     \includegraphics[width=8.0 cm]{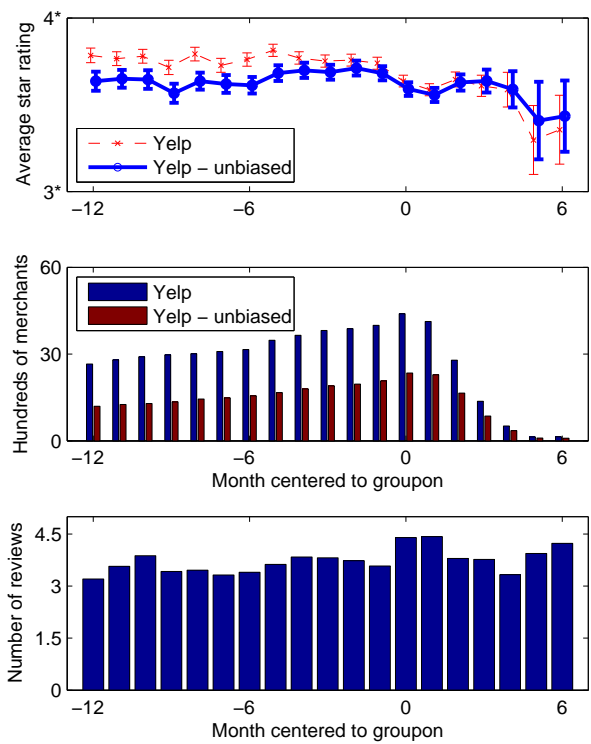}
 \end{center}
 \caption{{\small {\YelpUnbiased} dataset: Yelp Ratings by rating index.  The upper plot shows the mean of the average rating of every merchant for every month around the date that a merchant ran a Groupon deal (month 0) and the corresponding 95\% confidence intervals for {\YelpUnbiased} and {\Yelp}. The middle plot shows the number of merchants aggregated for each month. The lower plot shows the average number of reviews per merchant at each month. \label{fig:monthsPlotAggrMerchants}}}
\end{figure}

\section{Groupon and Yelp}
\label{sec:groupon}
In this Section we perform a case study on the effect that Groupon has on Yelp reviews in the light of the bias of the initial reviews. Our study shows that the reputational ramifications on monthly average reviews caused by Groupon are overestimated using {\Yelp}.

We begin our analysis by creating a dataset that suffers less from the bias of the initial reviews. As we argued above the bias is present in the first 20 reviews of each business. Therefore we remove the first 20 reviews for all merchants. This leads to a dataset with 3071 merchants and 170631 total reviews. We call the new dataset {\YelpUnbiased}. 

Next we reproduce the plot presented by Byers et al.~\cite{Byers2011} on the complete {\Yelp} dataset plus additional information about the data presented therein. The top plot of Figure~\ref{fig:yelp-biased-hist} illustrates the average of the average month rating of merchants after centering all reviews around their Groupon launch date. Every aggregate is the mean of the average rating per merchant at that month. We note that each month contains a different set of merchants. Line 1 is qualitatively similar to the one presented in~\cite{Byers2011}-Figure 9. Small differences stem from the fact that we have used a more recent snapshot of the dataset.   

The second subplot illustrates the percentage of ratings in the $\left[1, 20\right]$ index interval for each aggregate point in the top subplot. Observe that the second subplot correlates significantly with the first. The correlation coefficient between the points in the two subplots, i.e., the percentage of biased-records and the average monthly rating is as high as 0.87.

Indeed the number of ratings in the $\left[1, 20\right]$  interval is greater in the \emph{pre-Groupon} period.  Overall 34\% of the records in the pre-Groupon area are in the $\left[1, 20\right]$ index interval while 25\% of the post-Groupon area belong to the same interval. Indicatively we report that the percentage of records in the bias area in the dataset is 34\% for the point that corresponds to 6 months before the Groupon offer and 16\% (less than a half) for the point that corresponds to 6 months after Groupon. 

The third subplot shows the number of businesses present in each datapoint of the top plot. Clearly each month is \emph{not} equally represented in the dataset. We report that the 5th and 6th month are aggregates over 135 and 144 merchants respectively while the 0 month average is an aggregate over 3994 merchants. This explains the loose confidence intervals that span close to 0.4 stars.

After identifying the large correlation between the bias and the average rating in the complete {\Yelp} dataset we study Groupon's effect on Yelp reputation after we remove the bias. We redraw the plot using the {\YelpUnbiased} dataset in Figure~\ref{fig:monthsPlotAggrMerchants}. We also redraw the {\Yelp} line from Figure~\ref{fig:yelp-biased-hist} for comparison. Observe that the change in the reputation in the post-Groupon era is largely due to the initial review Yelp bias. We note that the confidence level of the points is  set to 95\%.  Once the bias is removed we observe that the average ratings in the pre-Groupon area were overestimated using {\Yelp}.   We report that the presence of the bias causes an \emph{increase} to the average rating  by 0.15 stars 6 months before the Groupon deal while it also causes the average rating to \emph{decrease} by 0.03 stars four months after the Groupon deal. 

We compute in the {\YelpUnbiased} dataset that the rating decreases by 0.04 stars after a Groupon deal in the 20-40 index and 0.05 overall. Thus, there seems to be a shift in the average rating once a business is exposed to a wide audience: an increase in the number of ratings by 80\% comes with a small decrease of 0.05 stars. Furthermore, comparing the overall number to 0.12 stars reported previously~\cite{Byers2011, mitbyers} we conclude that analyzing the biased dataset results to overestimating Groupon's reputational ramifications by a factor of 240\%.
 
We now focus on each month's average review in the post-Groupon era. The top subplot of Figure~\ref{fig:monthsPlotAggrMerchants} shows clearly that months 2,3, and 4 in the post-Groupon era do not exhibit any significant change compared to the entire pre-Groupon era. Furthermore, the middle subplot of Figure~\ref{fig:monthsPlotAggrMerchants} shows that months 5 and 6 contain less than one hundred merchants, a small sample of the merchants aggregated in the other points of the plot. The different mix of merchants and the loose confidence intervals cannot support conclusions for this area.
We now draw the average number of reviews per merchant for each month in the bottom subplot of Figure~\ref{fig:monthsPlotAggrMerchants}. Observe that the average monthly review anticorrelates with the number of reviews. The correlation between the average review and the number of reviews is -0.36. The correlation coefficient is -0.41 for the post-Groupon area. Assuming that the number of reviews is a proxy of the number of customers that the merchant receives, this anticorrelation suggests that larger traffic at a business may correspond to lower ratings. 

\smallskip
\noindent
\textbf{Limitations. } Various factors add noise to this analysis. Each aggregate point in Figures~\ref{fig:yelp-biased-hist} and~\ref{fig:monthsPlotAggrMerchants} is computed over a different set of merchants. Also, the mix of months in which each review was posted differs significantly among data points. Indicatively, more than 90\% of the reviews aggregated in month 0 were posted some time between January and August.

\section{Conclusion}
\label{sec:conclusion}
In this paper we identified and corrected for a warm-start bias that appears in Yelp data. In particular, we demonstrated that the first reviews that a business receives in Yelp significantly overestimate its eventual reputation. The explanation of this phenomenon may be that a business gets exposure to a limited and possibly favorable audience during its initial steps. It is worth noting that such a bias is not present when a merchant receives his first review after performing a Groupon deal. Presumably businesses that start their Yelp presence after they launch a Groupon deal are directly exposed to the broad audience and receive reviews that are more likely to be real.

We also performed a case study to explore the effect of Groupon deals on the monthly average Yelp ratings of a business.  We showed that the percentage of initial reviews in an average rating correlates significantly with the average rating. In other words, an uneven distribution of initial reviews introduces a bias to the analysis of reputational ramifications.  Therefore, we removed the bias and revisited the effect of Groupon on Yelp ratings to identify a subtle anticorrelation between the number of reviews a business receives in a month and its monthly average rating.

The analysis and methodology presented in this paper points to the importance of detecting and correcting biases present in online reputational systems such as Yelp. 

\section*{Acknowledgments}
The author would like to acknowledge Aristides Gionis and Rajesh Parekh for insightful comments on this study. The author would also like to acknowledge John Byers, Michael Mitzenmacher, and Georgios Zervas for their contributions in the data collection. 

\bibliographystyle{abbrv}
\bibliography{biblio}

\end{document}